\documentclass[prb,twocolumn,showpacs]{revtex4}\usepackage[T1]{fontenc}
\usepackage{lmodern}
\usepackage{graphicx}
\usepackage{epsfig}
\usepackage{bbm}

\begin{document}

\title{A Case Study of a Scientific Blunder  : History and Philosophical Teachings }

\author{P. Lederer \\ 
Directeur de Recherche honoraire au C.N.R.S.\\
14 rue du Cardinal Lemoine, 75005-Paris\\
e-mail  : pascal.lederer@u-psud.fr\\
Laboratoire de Physique des Solides\\
Universit\'e Paris-Saclay, France
33(0)662984051 }

\begin{abstract}
 In 1988, in cooperation with a team of experimental physicists, a Condensed Matter theorist, X,  published in Physical Review Letters a crucial experimental result dealing with a revolutionary new theory. The conclusions of the paper were proved incorrect a few months later. I discuss the various factors -- scientific, instrumental, but also psychological, sociological ones --  which led to this blunder.  I believe this story  sheds some light  on the process of scientific discovery, explanation, falsification,  confirmation, and errors. 

\end{abstract}
\maketitle
\newpage

\vspace{20mm}
\section{Introduction}

In 1986, a stunning experimental discovery rocked the world of physics, that of High Temperature Superconductivity \cite{bednorz}.  It was discovered in a material, $La-Ba-Cu-0$ (hereafter $LBCO$), for which no learned solid state physicist, either on the theoretical side or on the experimental one, would ever have believed it to occur. The superconductivity was  observed at temperatures which had been thought until then to be impossibly high, as compared to that of all studied superconducting  metals since the discovery of superconductivity in 1911 \cite{KO}.

The search for higher temperature superconductivity (the highest one until then was $21^0K$) had motivated some researchers in the previous years, most had given up. Superconductivity was considered widely among condensed matter physicists as a "dead" topic, as a Nobel laureate physicist declared,  a topic where nothing significantly new would ever be found any more.   The vast majority of researchers had moved to other fields.

Until then, it had been common knowledge among physicists that magnetic impurities were detrimental to superconductivity, as proved by scores of experiments (see for example ref.\cite{tinkham}). A small concentration of magnetic impurities in a superconductor was known, and understood, to lower drastically the temperature below which superconductivity was present. In the new superconducting material, however, magnetic atoms were not only present, but in fact densely so throughout the material (see for example ref.\cite{ledererHTS}). In fact $LBCO$  was very close to being an antiferromagnet, i.e. a material where magnetic ions are present in a regular crystaline array, with an alternate order in the lattice of $+$ spins and $-$ spins at low temperatures (in contrast with ferromagnets where all spins have the same direction at low temperatures). Not only are magnetic ions dense in the new superconducting material, but superconductivity  persists up to temperatures significantly larger than observed until then in non magnetic materials. In fact, very rapidly, in superconductors of a class similar to $LBCO$,  superconducting temperatures reached an order of magnitude larger than any previously known material.

It was obvious to the vast majority of physicists that they were dealing with a qualitatively new phenomenon, immediately dubbed "High Temperature Superconductivity"(hereafter HTS). It quickly made head titles in the news of the world, predictions of a new technological era were made, or even an industrial revolution was hypothesized   : in particular, if superconductivity could be made to persist up to ordinary temperatures,  the old industrial problem of electricity storage, so crucial for  attempts to fight the current climate change,  promised to be solved, since superconducting rings can store electrical currents  with vanishingly small resistive losses for very long times. Spectacular experiments showed levitating objects, or even people, above superconducting chunks of the new material, etc..

Technological expectations in various industries were huge. Financial ones as well.

Suddenly, condensed matter physicists were in the world news!

As a result, hundreds of experimentalists turned to investigate the new materials, searching for even higher superconducting temperatures, and hundreds of solid state theorists around the world  left more or less aside the projects they were previously interested in, to try and contribute  theoretical advances in the understanding of HTS. An intense world competition among individuals, groups, laboratories, universities and research institutes developed.

At the time, X, a Condensed Matter theorist  was at a turn of his career. He had led a small group of theorists during ten years with some success, about electronic properties of nearly one dimensional conductors. Those can be pictured as one dimensional atomic filaments weakly coupled by some chemical bonding, so that they exhibit very specific anisotropic properties. With his collaborators and  PhD students they had gained some recognition for the discovery and understanding of a new phenomenon, dubbed Field Induced Spin Density Waves \cite{ledererFISDW} (hereafter FISDW). The explanation  they had given of their observed Quantized Hall Effect had attracted attention in the physics world. Their theoretical tools were those of Quantum Field Theory, in a theoretical framework called "perturbation theory" to deal with interactions between electrons in solids. The latter is the convenient approach when there are sufficient hints that the phenomenon one wants to study cannot be understood if one neglects the interactions between the charge carriers (conduction electrons) within the material. Taking into account interactions between electrons -- contrary to what is needed to understand most physical properties of conductors such as Na, Cu, Ag, and a number of transition metals etc. -- in the FISDW phenomenon is mandatory. However the interaction energy scale (written symbollicaly $U$) in that system is small compared to the kinetic energy scale  of electrons, in other words the width (written $W$) of the electronic bands. This leads to what is called "perturbation theory"  : the theory relies on known results  when $U$ is zero, and introduces non zero $U$ as a small perturbation to the zero $U$ case. This method is useful, provided the limit for infinitely small $U$ is not singular. The theory of the "conventional" (low temperature) superconductivity is also  based on the notion that $U$ is very small compared to $W$ \cite{BCS}\footnote{In that case, the whole difficulty is that $U$ is negative, and perturbation theory fails.}, so was the vast majority of theoretical papers on the interacting electron gas in Condensed Matter physics until the discovery of HTS. This basic starting point in the former case is summarized by the inequality below  :
\begin{equation} \label{I}
 U<<W
\end{equation}
where both $U$ and $W$  are positive (repulsive interactions between electrons).

In fact, after writing his PhD in 1967 on magnetism in metals within this perturbation approach, which was the relevant one for magnetic transition metals, X became uneasy with the complications of perturbation expansions in powers of $U$ of higher and higher order, demanding more and more complicated diagrammatic methods. He decided  to explore the physics of systems governed by the opposite paradigm, the "strongly interacting limit", described by the following inequality  :
\begin{equation}\label{II}
U>>W
\end{equation}
The line of research based on equation (\ref{II}) requires completely different theoretical tools, largely undeveloped ones at the time, and appeared  to be  relevant  in a much more reduced number of experimental situations  : solid bcc $^3 He$, magnetic insulators such as Copper or Iron oxides (i.e. $Cu0$ or $Fe_20_3$) and a few others. The challenge was interesting, X learnt some different physics, published a few papers, but after 1973, the number of interesting experiments stayed quite reduced, the interesting physics of $^3He$ required  lower and lower temperatures, increasingly difficult to reach experimentally  : X decided eventually to revert to the study of experimental systems which were explored in his laboratory, and for which the approach of equation  (\ref{I}) was the relevant one. This led to the FISDW work mentioned above, which was completed and published in 1986.

\begin{figure}[t]\label{structure}
\centering
\includegraphics[width=10.5cm,angle=0]{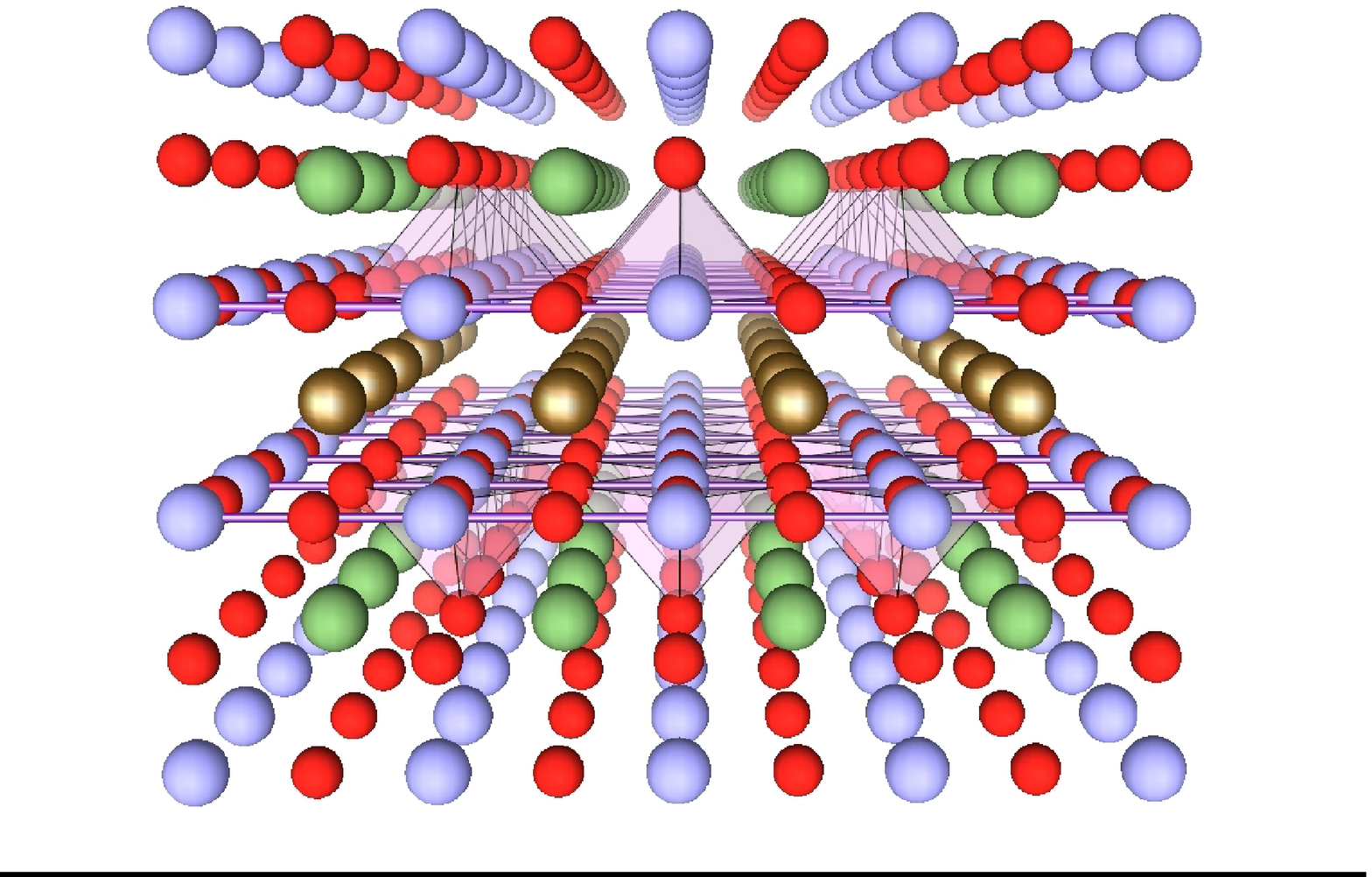}
\caption{\footnotesize{This figure exhibits the atomic structure of a typical HT Superconductor  : contrary to the vast majority of (lower temperature) BCS superconductors, which have a much simpler chemical structure and a much simpler (usually cubic) crystallographic one (in one guise or another)), this crystal is formed by parallel sheets of Cu-O  (Copper oxide) planes. The ``red'' atoms here are oxygen atoms, the ``blue'' ones are Cu (Copper)atoms;  two Cu-O planes  are separated by ``brown'' atoms (such as Ytrium or Lanthanum } while ``green'' atoms are divalent metals such as Ca or Ba. A large family of HT Superconductors exhibits the common feature of weakly coupled Cu-O planes which have a basic square elementary cell. The occurrence of linear arrays of Oxygen chains between copper planes is specific to the  particular  chemical (YBaCuO) shown above. This structure is a strong hint that the essential superconductivity properties of this material are connected with the Cu-O planes. This feature is present in a whole family of Cu-0 HTS compounds. I am indebted to Julien Bobroff for providing me with the file for this figure. }
\label{fig01}
\end{figure}

When HTS erupted, X was ready to dive into its mysteries, and perturbation theory based on (\ref{I}) seemed a sensible starting point. X put to use his previous experience on theory  based on this approach. The new material was an anisotropic one, formed of parallel 2D arrays of $CuO$ planes, weakly coupled to one another along the direction parallel to the plane, as depicted on  figure (\ref{fig01}).

A new theory of the interacting electron gas in two dimensions (2D) had just been published \cite{schulz} in a Renormalization Group approach within perturbation theory  about a model which was an idealisation of the $LBCO$ structure represented in figure (\ref{structure}). The first try by X at the explanation of HTS was to start from  this idealized model and to  take  into account a necessary improvement. The results were encouraging  : they confirmed that the improved model exhibited a tendency  to enhanced superconductivity. That work was sent for publication, and was published \cite{ledererlmp} in 1987.

Then erupted the revolutionary Resonating Valence Bond (hereafter RVB) theory of Philip Anderson and collaborators \cite{PWA}.

For the purpose of this paper, it is not essential to go into the details of the RVB proposal. Some however, are given in the Appendix, section (\ref{append}).

When X read the paper by Philip Anderson (\cite{PWA}), he was immediately deeply impressed by its audacity, and by the  novelty of the scheme he proposed to explain HTS. He had been a long time admirer of Phil Anderson's many brilliant and original contributions to physics, in various fields. The work on HTS in reference \cite{PWA} was based on the hypothesis that the basic theoretical framework relevant to HTS was that of the inequality (\ref{II}). Instead of starting from the properties of a weakly interacting, two dimensional, conducting electron gas, the starting point was that of the strongly interacting electron system of a (two dimensional) quantum magnetic insulator.

X, as a large number of physicists, was immediately seduced\footnote{This seduction is commented upon in a later section of this paper} by the RVB approach; there were various reasons for this  :  the audacity and the novelty of his proposal, the rationality of it, given Phil Anderson's previous work on RVB (see reference \cite{ledererHTS} for details), his scientific  prestige; those three factors combined  with the novelty of the HTS phenomenon, the novelty of its experimental features, which were utterly contrary to all scientific rationale until then.

X decided to give up his previous approach based on condition (\ref{I}) and decided to work on Phil Anderson's RVB approach, based on condition (\ref{II}). Among condensed matter physicists, a fierce battle developed between champions of the weak interaction approach (expression (\ref{I})) and those of the strong one (expression (\ref{II})), as described and analysed in ref.\cite{ledererHTS}. Both approaches seemed so contradictory to one another that two main lines of theoretical work developed which it seemed impossible tp reconcile. 

In August 1987, X flew with his family to Tokyo, where he was to spend one year as an invited lecturer, in Hidetoshi Fukuyama's ISSP Tokyo\footnote{ISSP $=$ Institute for Solid State Physics} group. Hidetoshi Fukuyama, a brilliant theorist,  had spent some time at Bell Labs, at the time of its world fame, in collaboration with Phil Anderson. He had published many brilliant papers with the latter, together, among others, with Maurice Rice, another physicist X admired. Both had immediately adopted Phil Anderson's views and were active in developing the RVB strong interaction scheme. 

\section{Invited Scientist in  Tokyo University}\label{invited}
X arrived in Tokyo and attended at first the International Conference on Superconductivity which was held that year in Japan.

A crucial\footnote{Crucial predictions and experiments are discussed in  section (\ref{episteme}).} prediction of the RVB proposal was the existence at low temperature in  superconducting HTS materials of a specific heat $C_v$ linear in temperature, i. e.
\begin{equation}\label{specific}
C_v = \gamma T
\end{equation}
where $\gamma$ in equation (\ref{specific}) is a constant for a given material. $\gamma$ was known to be related to the "density of states at the Fermi level", a well known object to condensed matter physicists. In all ordinary superconductors so far, theory and experiments agreed that  $\gamma$ was zero, in contrast to the normal (i.e. conducting but not superconducting) metal case where it is non zero,  proportional to a quantity characteristic of the metal (the density of states at the Fermi level). A non zero sizeable $\gamma$ in the new superconductors was a stunning prediction, a crucial one. That prediction was based on a revolutionary part of the RVB proposal : the separation of spin and charge degrees of freedom in a strongly interacting electron liquid, and the existence of a Fermi surface for neutral spin excitations in the new (HTS) superconducting state. The hypothesis was that strong interactions could lead to new emerging fundamental particles, the spinless charged particles called "holons" and the neutral spin particles called "spinons". The "bare" (isolated) electrons simply disappeared as elementary excitations in the strongly interacting liquid! This idea appeared attractive to X because he had predicted the existence of new particles called spin polarons in solid $3He$ \cite{ledererpolaron}, another case of strongly interacting elementary fermions.

 Accordingly X paid special attention during the Superconductivity Conference to all experimental reports on the specific heat of $LBCO$. There were large variations in the results; sometimes a non zero $\gamma$ term was observed, with various magnitudes, sometimes there were none. The amount of Ba (Baryum) impurities to obtain  superconductivity in $LBCO$ was not accurately reported nor controlled. The Conference, where disagreements among theorists about the  framework relevant to HTS were blatant, did not allow any conclusion to be drawn. The separation between theory schools based on (\ref{I}) or (\ref{II}) was total. Theorists belonging to the first school thought that those belonging to the second school were a sect with a guru. Theorists belonging to the second one thought that they were daring  revolutionaries, while those believing in the first approach were dull oldtimers prone to routine thinking.

X started  to work in the ISSP laboratory in two directions; one was about consequences of the strong correlations (expression (\ref{II})) on the motion of doped holes in a magnetic insulator. The other was to visit two laboratories to talk to experimentalists to encourage them to clarify the experimental situation about the $\gamma$ term. X urged to study it whith well prepared and  characterized  samples with accurately measured  Ba concentration, and measuring the $\gamma$ term both as a function of Ba concentration and of the corresponding variation of the superconducting instability  temperature.

X talked to one group in Hokkaido, in the north of Japan, and to another one in Hiroshima, in the south.

Three months after arriving in Tokyo, X visited both groups, one after the other, to examine with them what were their results. What was shown to him in Hokkaido did not allow clear conclusions; accuracy in the Ba concentration was insufficient. It was agreed they had to improve that. In Hiroshima, the situation was worse  : no significant results were available.

\section{The paper is published}\label{publi}

However, to X' surprise and delight, three weeks after visiting the Hokkaido group, he received a fax from that group (internet was almost non existant...), exhibiting a significant rise of $\gamma$ coinciding with the rise of  the superconducting temperature! X had an immediate rise in adrenalin,  and reported the results excitedly to Hidetoshi Fukuyama and his colleagues in ISSP. The news spread with the speed of lightning in Japanese labs. Phil Anderson's  RVB theory  was experimentally confirmed, along with its spectacular new hypothesis on the behaviour of strongly correlated electronic matter!

X told the Hokkaido group they had to consider writing a report for publication.

A surprising news reached X two days later  : ISSP colleagues told him informally that the head of the Hiroshima group was reporting new experimental results on $LBCO$, in a small physics meeting in Tokyo, of which X had not been aware. He claimed his group had obtained experimental confirmation of the significant $\gamma$ term in LBCO!

This was astonishing. Three weeks before, the Hiroshima group X had visited  had no result. They had communicated no new results since then. How could X understand?

X' understanding of japanese ways of relating was obviously quite poor. X had noticed that some ISSP scientists whom he crossed in the hallways or in the staircase would behave as if he was transparent. He had understood that, not knowing his position in the hierarchy of science, they did not know how low they had to bow when they met him. The way out was for them to see through him.

With the claim by the Hiroshima group leader that  his group  had  results similar to those of  the Hokkaido group, without however acknowledging the latter results, X could not but imagine that the Hiroshima group was going to short the Hokkaido group about a historical result which had to become a landmark in the history of physics. On the other hand, the Hiroshima claim proved without doubt that the Hokkaido experimental results were reliable.

When two different experimental groups, working independently in different labs in different parts of the world find the same results on the same object in the same conditions, the probability that the results are correct is very high. Science philosophers call that the intersubjectivity argument for truth.

There was no sign from the Hiroshima group leader that he would share a publication with the Hokkaido group.

It was only fair, in such circumstances, to ensure that the Hokkaido group publish their results as quickly as possible, so as not to loose their obvious priority  : this  is a kind of ethical aspect  widely shared in research. X thought it was normal to add his name to  those of the experimentalists authors of the paper.  The paper was sent for publication in Physical Review Letters, which was the most widely read worldwide for new developments in physics.

Shortly thereafter, X traveled to a Conference in Aspen (Colorado, USA) where he reported about his work\footnote{X had published in Journal of the Physical Society of Japan the first paper showing the scale of HTS in the RVB framework could not be the band width $W$ of holes, but the exchange energy $J$ among spinons. This paper, perhaps because it was published in Japan \cite{ledererjap} and with 8 months delay (!), is hardly ever quoted, even though X reported about it in Aspen.} and the experimental results of the Hokkaido group. Phil Anderson was a few yards away from X when he delivered his speech. X was proud to bring him an important piece of experimental evidence on the $\gamma T$ term in $LBCO$ in favour of his RVB approach.

The paper on the low temperature  specific heat  in the superconducting phase of $LBCO$ was published (reference \cite{ledererhok}) only two weeks after it was submitted fro publication. This must have been one one of the most rapidly published papers in the history of Physical Review Letters. Usually, papers are sent to referees, who may ponder for months about the validity of the results, send various criticisms to the authors, ask for some change, for checking the validity of such and such statement for citing more or different authors, etc.. No such thing happened in this case. One may surmise that Phil Anderson's prestige  may  have helped in convincing the editors of the journal of the interest of a speedy publication.

The Hiroshima group announced a communication to be delivered at a Conference in Interlaken (Switzeland) two months later, where they confirmed the linear specific heat in $LBCO$.

Barely two months after the paper with the Hokkaido group appeared, with X' signature, it was spring time in Japan, the whole Institute was going to celebrate "Hanami", to watch the magnificent cherry trees blossoming, close to ISSP, ...in a graveyard near by. 

On the very eve of that important social event, a report arrived about new experiments in a compound dubbed $BSCO$, of the same class as $LBCO$, with a much higher superconducting temperature, and much more reliable crystal composition than that of $LBCO$.

There was no sign of the $C_v$  term linear in temperature in the superconducting phase. It was clear that for one reason or other, the results displayed in \cite{ledererhok} where not related to superconductivity\footnote{The linear specific term reported in $LBCO$ is due in fact to a particular impurity  disorder effect in metal/oxide compounds similar to $LBCO$.}.

X' participation in the joyful Hanami event is one of the worst memories of his stay in Japan. Not only had he publicly put his name on a paper reporting incorrect results, not only had he induced a japanese group to publish incorrect conclusions on a dubious experimental result;  by specifying in the head of the paper that he was a member of the ISSP, X had put a stain on its international reputation.  X had lost face. In japanese terms he had dishonoured his samura\"{i}, he had to commit seppuku.

X did not commit seppuku, but he  almost felt physically the weight of the shame on his shoulders. This shame and this stain remained with him during the remaining months of his stay in Japan.

\section{Lessons and comments}

At first sight, why spend any time writing about this episode? It looks simply like an experimental and theoretical blunder, a rather common event after all.  Why not let it rest in the graveyard of the host of incorrect papers?

However, I believe it deserves some more discussion. The development of this story, its context, its causes,  its results, have some philosophical, sociological and epistemological interest. 

To begin with, what was wrong with the experimental results? When one looks at the graph for $\gamma(c_{Ba})$ published in the specific heat paper \cite{ledererhok}, where $c_{Ba}$ is the $Ba$ concentration in the $LBCO$ samples, one finds no indication of experimental error bars. The probability is that if the error bars were indicated, a straight line in function of $c_{Ba}$ would be compatible with the data, indicating a phenomenon independent of superconductivity, possibly  triggered by $Ba$ impurities\footnote{As is established now.} and not superconductivity. The experiment in Hokkaido did measure a significant low temperature $\gamma$ term coinciding with the superconducting phase, and which was negligible in the non superconducting one,  but the graph of the measurement was too inaccurate to suggest that the RVB prediction was confirmed. A well known fact is that the significance of any measurement depends on the upper and lower limits of its accuracy.

So how come X did not worry about such indications? How come he did not ask the Hokkaido group to evaluate the error bars prior to publication? He worried at first about this, but dismissed his worries when the Hiroshima group leader claimed he had the same crooked curve, with zero $\gamma(c_{Ba})$ until the critical $Ba$ concentration  for superconductivity to appear was reached. The intersubjective proof induced him in relaxing this otherwise professional demand. 

This leads to a second question  : why did the Hiroshima group leader, who was a well known scientist, a physics Professor  in the Hiroshima University, announce he had the same results as the Hokkaido group? In fact the odds are that he had no such results, let alone results identical to those, incorrect ones, of the Hokkaido team.

Japan has $88$ national  universities, $7$ of  which are "imperial universities", founded before world war two. Tokyo University, Kyoto University, and Hokkaido University, and others, are parts of this  University elite. When X happened to mention to japanese people that he was lecturer  in Tokyo University, he was immediately considered with a lot of respect. Hiroshima is a national University, founded in $1949$, not an "imperial" one. Has this played a role in the attempt by the group leader in Hiroshima to  claim equal success with Hokkaido in the experimental investigation of the sensational RVB theory? No one will probably ever know, but such human passions ("sad passions" discussed by Spinoza) in the science activity are frequent.  The excitement about the HTS phenomenon was shared by most searchers in that field, with its promise for world scientific recognition in case of success. There are many occurrences, in the last decades of physics research, of deliberately  fake reports in some of the best science reviews such as Nature, Science, Physical Review Letters, etc.. They are sometimes connected to an attempt to get approval of  grants from specific peer committees for research funds, sometimes inspired by a search for recognition. There are numerous cases of scientific thefts, when a searcher publishes under his name a result obtained by others\footnote{X had been a victim of such unethical conduct by well known colleagues at an early stage of his career.}. The social pressure among scientists to succeed, to be invited for talks in conferences, for stable positions, for promotions, for research funds, is producing those failures of scientific ethics. In the case of HTS in 1986, with the world spotlights on, and the excitement among physicists, all the elements for such ethical failures were enhanced\footnote{This paper is written during the Covid-19 pandemy. Perhaps the race and competition among epidemiologists and searchers in biology to find a treatment, or a vaccine,  will help understand the HTS situation in 1986...}. 

It was highly recommended for theorists in the home laboratory of X to visit experimental facilities where experiments were conducted which were connected to theoretical problems of interest. Examining the apparatus, discussing with experimentalists how they were conducting experiments and what were the relevant theoretical questions  : this was thought to be  part of the theorist's job. Among his japanese colleagues in ISSP, the visits of X to experimental groups to discuss relevant experiments about HTS and RVB were perhaps not viewed positively, although this was  openly expressed in only rare occasions, because of politeness ethics. Similarly, X realized after a time that asking questions and emitting critical scientific comments while attending a science talk was not a polite thing to do...Had X been impolite with his Hiroshima colleagues? 

The episode I am discussing here is certainly due to a combination of various factors, of various human subjectivities, of cultural differences. Triggering a competition between two groups, belonging to different institutions was not a scientific mistake, but it may have been a "cultural" blunder.

When X prompted two different experimental groups to work on the confirmation, or the falsification, of the RVB prediction, the competition he had unleashed  could have been fruitful if both competing groups, while working with their own ways and  methods,  had cooperated and exchanged informations about their mutual progress, or the difficulties they faced. Such had not been the case. An obvious lesson is that the competition, together with  scientific cooperation between the two groups would have spared them, and X, the disgrace of a faulty publication, as well as misconceptions in the physics community during a few months...

What emerges from this discussion is the manifold of subjective, emotional, cultural, sociological factors involved in the course of scientific investigations. Although some of the factors discussed above are specific to the country where the present story took place, I believe a range of such factors are always present, with varying degrees of importance, irrespective -- nowadays -- of time or country.

\section{Epistemology}\label{episteme}
Although the following  is not intimately connected with the episode of the incorrect specific heat paper \cite{ledererhok}, it is useful to set the stage with general comments on the scientific endeavour of exploring inanimate matter and conceptions surrounding it.

Before commenting on connections of the story of reference \cite{ledererhok} with various  contemporary debates in epistemology, I briefly sketch some issues on the process of knowledge of inanimate matter. 

Can science only achieve a description and explanation of phenomena without reaching truths about the object itself? Philosophy schools have confronted each other since Heraklite, Lucretius, Epicure, then Diderot or Marx, who were  materialists, while Socrates or  Plato, then Berkeley, Kant or Hegel  were idealists. Idealists, for example Plato or Berkeley,  may claim that  ideas are more real than things. Materialists claim that matter comes before ideas : the latter are fundamentally the result of material processes, be they individual or social ones. Berkeley \cite{berkeley} taught that the world is a mental construction  : only sensations are real. Kant \cite{kant} was an idealist who taught that experiments could gain some insight  on nature, but could not reach a true knowledge of the thing in itself ("das Ding in Sich"). Hegel \cite{hegel} argued that antinomies found by Kant were in fact ubiquitous in  connections or processes of knowledge about nature and developed the logic of dialectical contradictions.  Duhem \cite{duhem},  Mach \cite{mach} or Carnap \cite{carnap}, followed by many present day science philosophers, believed that the task of knowledge is to account for appearences. They claim that statements about the real essence of things is metaphysical. This philosophical current, called positivism,  which can be dubbed "empiricism", is described in detail in ref. \cite{empiricism}.
Empiricism is sometimes contrasted with realism \cite{realism}, following which truths may be obtained about things in nature. There are various versions of realism; some admit that God is real, and are a form of spiritualism. Materialism \cite{marx,lenin} insists that matter is what exists independent of human  consciousness. It existed before life appeared on Earth, so that the human mind emerged from inanimate matter in the course of a long historical biological process. The mind is a material  process, knowledge of processes in nature results from the reciprocal interplay between human minds in society and things (i.e. matter) which exist independently of the former.
\subsection{What truth?}
What do we mean by truths about nature? 
Na\"{i}ve realists believes that knowledge can reach   truths which are  exact images of real things, or  exact re-produc-tions of real processes. The latter  exist irrespective of the subject's consciousness. An exact image is such that it co\"{i}ncides exactly with the thing or the process.
A more elaborate version of realism emphasizes the difference between absolute truths which cannot ever be questioned, and relative truths, which depend on various conditions, parameters, and accuracy of experiments. The latter are historically evolving with the evolution of theories, experimental tools, and practical applications of the scientific progress.
Materialists believe that ideas about nature, either individual or social ones, originate from sensations due to external stimulations. Ideas are only potentially material. Marx \cite{marx} taught that ideas may  become material forces when they are shared by masses of people.

Physicists are often spontaneously realists in their professional activity when not explicitly positivists, or unconsciously materialists. Popper's ideas\cite{popper} have seduced many because they seem to co\"{i}ncide with their practice  : theories are interesting and meaningful if they can be tested ("falsified"' in popperian terms). They generally do not know about Lakatos' \cite{lakatos} criticisms of the falsification theory, although they spontaneously resort to  the Lakatos  "protective belt" in  case a  theoretical prediction is shown to be experimentally wrong : this is exactly what happened in the RVB case I am discussing. They are used to consider that experiments have limits of accuracy, and that  a theoretical explanation is only valid as far as certain minor causal chains can be safely disregarded when accounting for the main effect they  are interested in. They usually believe that science allows to establish truths about matter. Some are undisputable, and may be called absolute; some are simultaneously absolute and relative  : indeed, their validity cannot be questionned,  once  conditions (pressure, temperature, magnetic field, etc.), and accuracy limits of measurements are specified. For instance, it is undisputable that $LBCO$ becomes superconducting in such and such conditions of temperature, pressure etc. The problem this paper discusses was the explanation for superconductivity to occur in this material, at quite larger temperatures than previously thought possible. Truths can be improved,  or made more precise when technology offers new tools for better or different investigations. The existence of Saturn moons that Galilee discovered with a rather primitive means of observation cannot be questioned, but astronomers know many more things about them, thanks to telescopes in satellites, etc.. 

The development of physics, i.e. of human knowledge about inanimate nature, is  a historical and social process. Since the end of the nineteenth century, more than ever before,  it  combines the subjective activity of individuals, theorists and/or experimentalists, with that of  groups of individuals, laboratories, networks of groups or laboratories, conferences where new results are publicized and where  critical  confrontations between conflicting views occur, scientific journals,  national funding policies, public opinions etc.. Scientific results play an increasing role in industrial processes and social life, a phenomenon which is also a practical proof of the effectiveness of science to re-produce \cite{seve}, in an approximate or exact fashion, a large number of actual real processes of nature.
 
A collective (social) physicist is formed as a result. This social activity deals with the subjective  process of knowledge of  objective properties and processes of matter. The latter can be studied experimentally with repeated experiments by different independent workers in different places under identical conditions. The possibility of repeated experiments under identical conditions by very different workers is one of the specific features of physics, chemistry or biology  research, as compared to many issues in social sciences, for example, where the course of history often prohibits repeating identical situations\footnote{Repetitive phenomena are known to occur in society, such as, in economics, fluctuations of prices \cite{marx}.}. It allows to reduce the subjective part of the knowledge process.  It allows to correct errors, as X learnt at his own cost \cite{ledererhok}, and to elaborate  grains of truth on inanimate matter, truths about the laws of nature, some of which may be absolute, undisputable, some  which are relative to a given level of technological progress, as discussed above.

As shown by the story told in section \ref{invited}, the intersubjective agreement, which is generally considered as a proof of reliable results, may also lead to errors due to contingent phenomena, such as quest for social recognition, competition between groups, etc.. When X heard that the Hiroshima group reported results analogous to those of the Hokka\"ido group , he did not deem it necessary to improve on error bars...At the same time, the later understanding that the results the Hokka\"ido group published \cite{ledererhok} were unreliable also relied on  intersubjective agreement. 

Thanks to technological advances, and to continuous renewed social needs, the knowledge process of nature  never stops asking new questions, abandoning sometimes refuted theories, and improving on previous approximate  re-productions of real processes with better and better accuracy.

Although I share with Popper \cite{popper1} the observation that there is a cumulative progress of science, I do not agree with him that all knowledge is provisional. According to him, a theory is only valid as long as it is falsifiable.  In my view, the continuous improvement of experimental techniques, the continuous improvement of experimental accuracies allows undisputable truths to emerge in various areas of physics, at a certain level of experimental accuracy.

 Among many examples, the history of superconductivity is a good counter example. The BCS theory on the superconductivity of superconducting simple metals\footnote{Such as Tin, Mercury, Zinc, Vanadium etc..} has undisputable results. The discovery of HTS in $LBCO$ and other cuprates showed that those  previously unexplored materials exhibit superconducting properties which  had not been suspected, but it did not lead to correct the theory of superconductivity in simple metals.  Whether  RVB theory will  end up being recognized as a fundamental addition to BCS theory for the understanding of HTS remains to be seen, as will be discussed later.

\subsection{General comments}

What I have described in the Introduction, namely the discovery of HTS, is a good example of what Bachelard \cite{bachelard} called a social production of science. $LBCO$ and all the High Temperature Superconductors of the same class, i.e. based on coupled copper oxide layers, are  chemically engineered compounds. They were made possible by advances in oxyde chemistry, both experimental and theoretical, by hosts of scientists of different fields. In that sense, HTS, as most advances of contemporary physics, is a social production. However, contrary to Bachelard's ambiguities about the subjectivity associated with social productions, HTS is also definitely an objective property of matter, in given conditions of temperature, pressure, magnetic field, independent of human thought as evidenced by its practical uses ( magnetic fields detection, levitating trains on superconducting material, permanent currents producing magnetic fields, for example), as well as the  infinitely many repeated experiments in hundreds of different laboratories in the world.

 The theorist\footnote{A "problem solver" as Popper put it\cite{popper1}.} interested in solving the mechanisms of HTS does not, as Althusser wrote \cite{althusser} when he battled with empiricism, work on the real object. The image of the  $LBCO$ crystal, which is shown on figure (\ref{structure}), is not the real object; it is a man-created image. The searchers work on an object of knowledge already elaborated by various social processes as discussed in the introduction. However they also work on a phenomenon which is independent of any thought process, i.e. superconductivity, an objective state of matter, however engineered by human practice  : the "theoretical practice" is not disconnected from the real object, it is intimately connected, through a complex sensible, historical, technological, ideological process, to the phenomenon displayed by the real object, as discussed by S\`eve \cite{seve}. X would have been happy to confirm the $\gamma$T term in the specific heat of HTS, but nature denied it, irrespective of his intentions and desires.

 I do not feel it necessary to discuss various philosophical trends (see for example reference \cite{stanford1} on skepticism) which question how one can be sure (have a "true belief") that superconductivity exists and has such and such properties. I share Hacking's \cite{hacking} argument on that matter  : hundred years after Marx \cite{marx}, he rediscovered the criterion of practice as a criterion of reality.

\subsection{Scientific revolutions.}

I have used in the Introduction the notion of scientific revolution when discussing about HTS and RVB. Scientific Revolutions have been discussed in particular by Bachelard \cite{bachelard} and  Kuhn \cite{Kuhn}. Based on the history of physics ( Newtonian physics, relativity, quantum mechanics, and others), the latter author distinguished normal phases of scientific activity, which end when the paradigms of this activity are replaced, after a period of crisis, by new paradigms. An example he discussed is the replacement of geocentrism by heliocentrism; the emergence of quantum mechanics could be quoted as another example, although it did not invalidate many results of classical mechanics in their own domain of validity for most macroscopic objects.

HTS has a number of aspects of a scientific revolution  : the unexpected material where it was discovered, the stunning increase of superconductivity critical temperature in the new material, the revolutionary theoretical proposal such as RVB\footnote{ Not to speak about other proposals such as the "Spin Bag", or the anyon theory, which are not discussed in this paper, but in ref.\cite{ledererHTS}}. Seventyfive  years after its discovery \cite{KO} in 1911, the belief that no superconductivity temperature larger than about 23$K$ would ever be observed was shared by the vast majority of searchers in the field. Until 1986, superconductivity  was thought to have no future as a research program. It was not a period of "normal science" in Kuhn's sense  : it was almost inactive. From 1986 on, the need for an explanation of HTS seemed to point to a need for a new theory, fundamentally different from the "old" BCS superconductors \cite{BCS}, as described  in ref.\cite{ledererHTS}. The RVB paradigm \cite{PWA}, at the basis of the episode described in this paper, might be one such new paradigm.

There is however a major difference with Kuhn's description of Scientific Revolutions. HTS did not falsify the BCS paradigm for the well known "old" superconductors. It opened a new field of condensed matter physics, about a new category of superconducting compounds.  As discussed in a later section, it may evolve into blending the modern theories with the BCS one, i.e. into blending paradigms based together on seemingly incompatible expression (\ref{I}) and (\ref{II}).

The category of Scientific Revolution  may include scientific revolutions which are based on new paradigms without falsifying altogether previous ones. 

Other comments deal with the reasons that led X to abandon the scheme based on (\ref{I}) and adopt as research program the scheme based entirely on (\ref{II}). 

There were scientific, esthetical, institutional, psychological reasons, and also unconscious ones which X understood only years later, all more or less intermingled.

\subsection{Scientific reasons} X believed that the novelty, audacity of the RVB proposal by Phil Anderson and collaborators \cite{PWA} corresponded to the amount of novelty and surprises associated with the stunning discovery of HTS in compounds where all previous superconductivity culture would have denied its possibility. X believed this offered a possibility for a new field of condensed matter science to develop, a field he had abandoned  ten years before for lack of experimental support. He believed that this new field -- strong interactions in Condensed Matter physics -- would develop within the non relativistic quantum theory which was the relevant theoretical framework to explain HTS. Thus his scientific reasons were both based on a number of beliefs  : true beliefs about quantum mechanics, crystal structure of $LBCO$, existence of superconductivity, etc., and a subjective belief on the validity of RVB.

\subsection{Esthetical reasons} X felt the RVB proposal was so new, so seemingly well adapted to the novel structure and chemical composition of $LBCO$ -- as compared to "old" BCS superconductors --  that there was an intellectual beauty about it. In 1973, Anderson \cite{PWA2} had developed the concept of RVB for  a novel ground state of an insulating linear array of quantum 1/2 spins \footnote{In fact a model somewhat analogous to a railroad, he called "railroad threstle"}; this idea was in turn a generalization, for an infinite array of spins,  of Linus Pauling's  theory for the benzene molecule. At the time, in 1973,  the idea was original and interesting, it was a new concept in the field of magnetic insulators, but it had no obvious experimental counterpart. What Anderson did in 1986 for the theory of HTS was to extend this theory to a planar array of quantum 1/2 spins, and to postulate that the injection of "holes" (equivalent to suppression of electrons in the $CuO$ plane) would result in a charged bosonic superfluid of charged spinless particles (dubbed "holons"), i.e. a superconductor\footnote{See the Appendix \ref{append} for more technical details}. The progression of theoretical ideas from the benzene molecule to the doped $CuO$ plane of spin 1/2 particles had for X a beauty in itself.

\subsection{Institutional reasons}  X was a  CNRS searcher in 1986. His very social existence was based on developing new theories, predicting new phenomena, investigating new condensed matter phenomena, explaining them, and publishing papers in good quality physics journals, so that the new published results would influence other research programs. He thought that as such, he had to participate in the attempt to understand, explain and develop the theory of HTS, both as a scientific challenge, and as an industrial one; HTS had potential applications for the fundamental problem of electricity storage, which would be an industrial revolution. If X made a  significant contribution to some  research topic, he  would gain recognition among peers, perhaps a promotion, perhaps fame, etc.. 

\subsection{Psychological reasons} Working in this new field, with its scientific and industrial challenges, was exciting; X had learned to value the scientific stimulation connected to working on a topic of world wide scientific interest  : research in such a field gave an impression of more intense intellectual life, together with the risk to fail in bringing forward significant novel results. 

X shared a very general desire among researchers for social recognition through scientific achievements.

I believe now X had also unconscious reasons  : he had met Phil Anderson several times in scientific meetings, during his career. Phil Anderson was seventeen years older, a scientific leader when X started research, and he had treated X in a friendly manner, showing appreciation for his PhD work... and his skills as a chess player. On the other hand, X resented the lack of support, scientific or moral, or human warmth, from his scientific adviser. He felt his life as a searcher would have been richer if Phil Anderson had been his PhD adviser. In other words, the latter was a benevolent father image. By immediately adopting his RVB proposal, X became symbolically part of his family. This included a number of his collaborators, whose work X admired. One of them was Hidetoshi Fukuyama, head of the theory group in ISSP.

Ever since Plato and Aristotle, philosophers have discussed about the reasons to act  : in the present case, I tried to describe the various reasons which led X to work within a scheme based on relation (\ref{II}) rather than (\ref{I}). There are rich debates among them about the differences between "normative reasons" and "motivating", or "explanatory" ones. A normative reason is a reason to act. A motivating reason is the reason for which one does something. In the case of X, both types were clearly entangled. A study and complete bibliography about this topic is found in reference \cite{alvarez}. 

\subsection{Anarchist epistemology?}
The various personal reasons X had to choose one paradigm rather than another, as I have discussed above, seem good examples of what Feyerabend discussed in his book {\it{Against Method}} \cite{feyer,feyer2}. He criticizes the notion of  scientific activity as  based on a universal rational method. He stresses subjective reasons to adopt a theory. Although X seems to offer a good  example in support of his claims, the reasons of X were not devoid of belief in a fairly general method of science  : he took into account new experimental results, which, in view of existing wisdom about superconductivity, seemed to require new concepts, and to confront them with experiments.

 Phil Anderson's RVB theory was a scientifically attractive proposal. X decided to study it theoretically, and simultaneously to test it experimentally, with the help of skilled experimental physicists. Furthermore, a large number of Condensed Matter physicists engaged in the same research program on RVB. Many had probably reasons analogous to the scientific ones of X, perhaps a sizeable number shared his esthetic reasons, perhaps others also tended, as X did, to pay attention to RVB because of Phil Anderson's fame as a physicist. Very few had psychological or unconscious  reasons analogous to those of X. For the vast majority, what mattered was to develop an understanding of HTS and of its phenomena. Individual reasons had in the end little or no weight in the social process of science.

I have described elsewhere (ref.\cite{ledererHTS}) how various other theoretical ideas, different from RVB, were proposed since 1986  by various scientific leaders to explain HTS. I emphasized that each such new proposal was rooted in each author's scientific past, together with a general undisputable background knowledge of Condensed Matter physics. This reminds  in some sense of Feyerabend's remark on the weight of existing research programs to prevent new ones from developing. It also seems to support Feyerabend's relativism  : various groups of scientists believed in various theories, some of which were incommensurable  : at first sight  expressions (\ref{I}) and (\ref{II}) are mutually exclusive. This in turn looks like an example for which the Duhem-Quine thesis  \cite{duhem,quine} on the underdetermination of theory by experiments is temporarily valid. However, this state of scientific anarchy is but a transitory aspect of science progress, the process of research is constrained by the objective properties of inanimate matter. The latter allow  to correct scientific programs with a sufficient "protective belt" (see Lakatos \cite{lakatos}), discriminate between theories  after  a certain time. For instance Duhem's energeticism \cite{duhem} eventually had to be abandoned in favour of Boltzman's atomic theories. In HTS, there still prevails, at the time of this writing,  a state somewhat analogous to Feyerabend's anarchist theory of science or to some of his  relativism. However, in spite of the many persisting  disagreements among researchers on HTS, all agree on a number of basic physical laws governing superconductivity, which are undisputable properties of matter. 

 \subsection{Establishing truths} The absence of a $\gamma$ term due to supercondcutivity in HTS cuprates is now an  irreversible truth. This is now established by various careful measurements by different searchers. Duhem \cite{duhem}, following Kant's "Ding an sich",  claimed  that the role of theory is to account for appearances, and that no certainty can be obtained on the ontology of things causing phenomena. If instruments are involved in an experiment, they can lead to experimental errors, they may have defects, and lead to artefacts. At first sight, the Hokkaido paper \cite{ledererhok} would be a good proof of that thesis. But  it is  valid at a very restricted level, at most, for a certain time, for a few individuals or groups of individuals, and has no lasting universal validity, as discussed in ref. \cite{lederer3}. As suggested by the Hiroshima part of this story, other sources of errors, due to human passions involved in scientific research may play a role during a certain time. But  scientific activity is a social one, as stressed above. Based on  various well established technologies such as calorimetry, chemical synthesis of copper oxyde compounds, etc., contributions from various research groups in the world rapidly corrected the faulty paper and established the truth about the non existence of a $\gamma$ term in the superconducting phase of HTS. 

\subsection{ Crucial experiment. Popper's falsificationism falsified}

Bacon \cite{bacon} has introduced the notion of crucial experiment, which allows to discriminate between a truthful theory and an erroneous one. Duhem \cite{duhem} has argued that there is no such thing. I have argued elsewhere \cite{lederer3} that Duhem's stand is falsified by his own  admittance that "{\it there is certainty about the theory of vibrating strings}". The notion of crucial experiment is vindicated in many cases by scientific practice and is intimately connected with the notion of truth, as I discussed in a previous paper \cite{lederer}. The absence of a $\gamma$ term in the specific heat of $LBCO$ was a crucial result, but only in a limited way, as discussed below.

Following Popper \cite{popper}, the RVB proposal on HTS is a bona fide scientific theory, since it can be falsified. Phil Anderson's RVB original proposal predicted a specific heat linear in temperature in the superconductive phase. It is now well established that there is no such phenomenon in HTS.

According to Popper, this would ensure the falsification of RVB theory. It should be abandoned. However, this did not  happen. The theory was slightly modified to take into account the zero $\gamma$ result  : instead of proposing a spherical symmetry ("s-wave" symmetry) for the superconducting order parameter in HTS,  a so-called "d-wave" symmetry was introduced\footnote{see ref.\cite{tinkham}}. With this symmetry, the theory does not predict a finite $\gamma$ term any more  in the superconducting phase. This was not simply an ad-hoc change  : it was more coherent with the basic hypothesis of inequality (\ref{II}). Indeed, a superconducting order parameter with d-wave symmetry implies that the interaction energy $U$ between electrons is minimized in comparison with the s-wave case. Thus the stability of the superconducting phase is enhanced compared to the "s-wave" proposal.  The absence of a $\gamma$ term in he specific heat of  HTS is crucial to eliminate the possibility of s-wave symmetry for the HTS order parameter, not for the validity of RVB.

This is in line with Lakatos' \cite{lakatos} and Feyerabend's \cite{feyer} criticisms of Popper's Demarcation Criterion. Lakatos argues that a scientific research program consists of a "hard core" and a protective belt. Outside the hard core, there are a variety of auxiliary hypothesis to protect the hard core; changes and adjustments may occur in the protective belt, leaving the hard core untouched. This is what happened to the RVB research program when it was proved that the $\gamma$ term is absent in HTS, at the cost of changing the s-wave original proposal to the d-wave one.

Following Feyerabend, conformism is likely among scientists to favour old theories rather than revolutionary ones. The history of HTS offers both examples and counter examples, so that Lakatos seems to be correct in pointing out the role of conflicts between theories in the progress of science. Another way (see section (\ref{contra})) of stating this fact is that contradictions between subjectivities in theory, which reflect the dialecticity of nature, may be superseded by the development of science.

Popper, Lakatos, and Feyerabend all neglect the role of practice to prove at least in an approximate way various truths about nature. "Old" superconductivity and HTS seem to require different theories, but both account, even if approximately, for an undisputable fact of nature, i.e. superconductivity  : they re-produce all major observed processes due to superconductivity. The latter are put to use in various industrial applications.

In contrast with relativistic Feyerabend's positions \cite{feyer2}, no serious physicist is free nowadays to argue that a non zero $\gamma$  exists in HTS compounds of the $LBCO$ class, because the incorrect paper \cite{ledererhok} has undisputedly been shown incorrect. 

The conclusion of this paragraph is that RVB continued to be an active research program, even though one initial prediction was proved wrong. To this day, first half of year 2021, 35 years after the HTS discovery, the battle among theorists about the mechanism of HTS is still raging (see for more details ref. (\cite{ledererHTS}).

\subsection{ Contradictions?}\label{contra}
I mentioned in the Introduction that X' first published theory paper on  HTS \cite{ledererlmp} was  based on the weak coupling hypothesis (inequality (\ref{I})). This paper was written before X left for his one year stay in Japan. He reported about this paper only once in an international conference (in Genova in 1986). This report did not attract attention at the time. Thereafter X became  so convinced that the RVB picture based on inequality (\ref{II}) was the correct framework that he never even gave another talk about his weak coupling  work in Japan. X felt it was an irrelevant approach, it would not interest his colleagues. His host, Prof. Hidetoshi Fukuyama  was working on the strong coupling RVB theory : it would have been a loss of time to spend time reporting work on an approach both X and his host thought senseless.

Both approaches seemed incompatible. If one thought the weak coupling  approach (\ref{I}) was the right one, no attention  would be payed to works based on the strong coupling approach (\ref{II}), and vice versa.  Groups working on the  hypothesis (\ref{I}) would not talk to groups working on (\ref{II}) and vice versa. The development of this battle  is described in  more details in reference \cite{ledererHTS}.

Inequalities (\ref{I}) and (\ref{II}) are a particular example of a general couple of contrary conditions which are basic in the description of physical processes  : (\ref{I}) is a special case of the domination of kinetic energy over potential energy, contrary to (\ref{II}). A paradigmatic example is the classical harmonic oscillator. Depending on which term dominates the total energy, the phase oscillates periodically around zero, or increases periodically by $2\pi$. In both cases, both energies transform in time one into another. Internal energy and entropy are another couple of contraries in the free energy  : the internal energy drives order, the entropy drives disorder.

There are well known examples of metals where inequality (\ref{I}) applies without discussion in the description of their metallic properties :  $Au, Ag, Cu, $, transition elements of the first series in the periodic table,  etc.. There are well known examples of magnetic insulators where inequality (\ref{II}) applies without discussion  : $CuO, Fe_2O_3$, and all so called Mott insulators, or Quantum Hall systems \cite{lederer} . Is it conceivable that materials exist -- such as perhaps HTS materials -- the theory of which has to resort simultaneously to both inequalities?

Sixteen years after X' weak coupling paper on HTS \cite{ledererlmp} was published, the developments of experimental work on HTS caused theory to evolve towards a combination of (strong coupling) RVB theory (expression (\ref{II})) and weak coupling theory (expression (\ref{I}));  the results of the 1986 paper by X et al.  were rediscovered. Citations of that paper started to appear in HTS research theory papers by groups who worked in developing the RVB approach.

In other words, research on HTS evolved after almost two decades towards a simultaneous account of both  (contradictory) inequalities (\ref{I}) and (\ref{II})\footnote{This is apparent in particular in the work by the ETH Zurich group led by Maurice Rice, another historical collaborator of Phil Anderson's}.

How could the theory evolve in such a way? Along the years, a tremendous amount of different experimental techniques have been used to study superconducting cuprates : thermal measurements (heat capacity, thermal conductivity, thermoelectricity), conductivity measurements (ac, dc, with a large range of frequencies), diffraction measurements ( neutrons, X-rays, electrons), measurements under magnetic fields in a wide range of temperatures, muon annihilation measurements, tunneling measurements (with the field vector in various directions relative to the HTS crystal symmetry axis) , NMR measurements, sound velocity measurements, Josephson junctions, and yet others I forget about, with all the possible crossed experiments using simultaneously some of the the techniques mentioned above. It turns out that some phenomena seem to favour a strong coupling theory, such as the famous "spin gap" in the spin susceptibility at temperatures above the superconductivity one, or the spin glass transition; others point out to a weak coupling one, such as the observation of a Fermi surface compatible with BCS d-wave superconductivity or the observation of  Charge Density Waves \cite{frachet,michon}.

It is becoming increasingly clear these days, that in the High Temperature Superconductors, various other symmetry breaking instabilities compete with superconductivity : Charge Density Waves break the crystal symmetry, Spin Density Waves, or Antiferromagnetism, add a breaking of time reversal symmetry , and perhaps other other broken symmetries, such as commensurate/incommensurate transitions. The multiple contradictions linked with such instabilities, not to mention the order/disorder competition between Free energy and Entropy, may well be an intrinsic feature of the eventual satisfactory theoretical re-production of HTS. These experimental facts call for a development of dialectical materialism to understand the processes which involve multipolar connected sets of contradictions\footnote{Such theoretical developments are perhaps what is needed to explore the philosophical content of the currently popular theme of "complexity.}, instead of the standard bipolar contradiction discussed in most books about marxism. I have already mentioned the need for such philosophical  developments dictated by HTS experimental findings and theoretical difficulties in ref. \cite{ledererHTS}.

There are various examples, in the history of physics, of successful theories which have superseded what had appeared during many years as contradictory, i.e. incompatible ones. Quantum physics for example supersedes the theoretical contradictions between  the continuous (wave like) and the discrete (corpuscular) theories of light or microscopic particles. The theory of ferromagnetism developed during years along two seemingly incompatible lines  : that based on localized electrons on atomic sites in a crystal, and that based on  band theory i.e. on  electronic wave functions extending over the whole crystal volume. It is now based on a picture with both extended wave functions and localized magnetic moments (see the discussion in ref.\cite{lederermills}). Another example is the long lasting historical battle between energeticists \`a la Duhem and exponents (Boltzman) of the atomic theory of matter.

Even though  X' first HTS theoretical paper had results showing  that the "weak coupling" theory based on (\ref{I}) did exhibit interesting  features (for example a sharp rise in the superconductivity  critical temperature upon doping),  in favour of framework (\ref{I}), he neglected this result altogether after he became convinced of the relevance of the strong correlation scheme based on (\ref{II}). 

This attitude of X was a dichotomic one  : either (\ref{I}) had to be the valid starting point, or (\ref{II}). After RVB theory appeared, X became blind to the interesting results of his own work! He had been probably  too ignorant of Hegel's, Marx' and Engels' philosophical work on dialectics \cite{marx,engels,hegel}.  Had he been more learned in philosophy, X might  had been capable of  thinking that theories based on such dichotomies as the opposition between (\ref{I}) and (\ref{II}) often account, each,  for only a partial aspect of the processes at work in the real object. Theory is often, if not always, constrained to evolve in time to take into account contradictory aspects of things, as forced upon knowledge by the growth of experimental information and theory confrontations. Theory evolves in time to embody and supersede contradictions in the re-production \cite{seve} of real processes. As far as HTS is concerned, data from a variety of experimental techniques seem to reveal that weak coupling and strong coupling theories are both simultaneously relevant. This is probably a sign that in the HTS material the actual situation is  :
\begin{equation}\label{III}
U \simeq W
\end{equation}

In some sense, this paper is a description of two scientific errors  : one is the publication of an incorrect experimental paper \cite{ledererhok}. The other is the failure by X to believe in the interest of his own work \cite{ledererlmp} in the search for a theoretical understanding of the HTS phenomenon.


 Had he not been dogmatic in his belief in (\ref{II}),  X would have  publicised among HTS researchers the interesting results obtained (\cite{schulz,ledererlmp}), on the basis of expression (\ref{I}), early in the HTS research development. This would have helped in promoting more quickly a better theory of HTS.

When a theory embodies contradictions in the representation of reality, can one argue that it is incoherent, only due the subjectivity of a human construct? Is Quantum Mechanics incoherent because it reconciles  corpuscular and  wave-like phenomena of microscopic matter? This would deny that experiments yield results which are, at least in parts, independent of the preconceptions of theorists or experimentalists. Even if experiments are laden with theoretical preconceptions, the phenomena displayed by those experiments originate from the thing under study, irrespective of what humans may wish or think. The idea that contradictions are an ontological aspects of things in nature has been developed in particular by Engels \cite{engels}, based on Hegel's dialectics \cite{hegel}. This is considered by many philosophers of science as nonsensical\cite{feyer,feyer2,lakatos,popper}. What can be stated safely at least is that theories are almost always driven to deal with conflicting theoretical terms. Physics continuously explains motions, evolutions, transitions, spontaneous breaking of symmetries, etc., on the ground of opposition between conflicting terms in the theory, as discussed above. The notion that conflicting terms in theory re-produce real ontological contraries is a logical one\cite{marx,engels,hegel}. My own opinion on that matter agrees with S\`eve's views \cite{seve} who argues about the "dialecticity" of nature. He means by that term that the connections and  processes of nature constrain  the theory to account for them in terms of dialectical developments of contradictions.

What emerges from the story told in this paper is that the process of knowledge of nature evolves in a network of various types of contradictions : 
\begin{itemize}
\item Individual subjective conflicts between the rational quest  for understanding phenomena and "sad passions": the quest for recognition or fame, for a good career, for a promotion, the psychological tendency to believe in a theory because of their author's scientific fame or personality etc..
\item Subjective conflicts between different actors of the knowledge process : different experimental schools, even though they may be complementary, will claim more scientific attention to their techniques and their results to obtain more funds at the expense of  others. Different theory schools led by different school leaders will fight each other, each claiming to have the only correct approach, each paying more attention to the others' flaws than to their own. This leads to the emergence of chapels who tend to relate in terms of antagonisms.
\item Those subjective individual and social conflicts originate from the contradictory phenomena which the investigated thing produces upon being observed and probed under different conditions or different tools : in the HTS research, sometimes they favour approach (\ref{I}), sometimes they favour (\ref{II}); sometimes they suggest objective connections to other types of contradictions; in the HTS physics  superconductivity and other symmetry breaking phenomena seem to coexist and/or oppose each other. Their connexions might also be  of symbiotic essence.
\item cultural subjective contradictions rooted in the dialecticity of nature. The century long battle between the corpuscular theory of light (Newton, Descartes) and the wave like theory (Huyghens,Fresnel) was based in the classical aristotelian belief that contradictions cannot exist, either in theory or in nature. In the HTS story, faced with contradictory theories such as RVB theory based on (\ref{II}) or BCS theory based on (\ref{I}) the notion that both cannot have  simultaneous validity  retreats with difficulty in spite of experimental  evidence. 
\item Large scale social contradictions are at work nowadays between  three main poles; on one hand dominant social forces who want knowledge to be put at use to benefit investors in the battle between large corporations  for the largest profit rates, on the other hand scientific workers who need more job stability and more funds to develop knowledge, independent of industrial application prospects; then society at large which wants scientific development to improve living conditions for the many.
\end{itemize}
 Two comments are in order:
\begin{itemize}
\item From the philosophical point of view, the HTS story  suggests an insufficient development of dialectics  to deal with  processes and relationships within nature which are  governed by a network of interrelated contradictions. The development of research on HTS only strengthens the point made in reference \cite{ledererHTS}.
\item The quest of humanity for the understanding of nature and its processes is replete with examples of beliefs which have lasted some time and have been proved wrong in time by confrontation of theory with experiments and practice. Some beliefs have lasted centuries, such as geocentrism; some have lasted decades, such as the belief in ether; the scientific blunder related in this paper has fed beliefs among physicists in a revolutionary finite $\gamma$ term in the superconducting phase of LSCO for a few weeks;   this belief has been proved wrong subsequently by better and different experiments by different groups. The social process of knowledge confronted with reality has overwhelmed the intricate network of contradictions discussed above to reach a lasting truth. 
\end{itemize}

\section{Conclusion}
I have described how X was led to put his name on an experimental paper the results and conclusions of which were found a few months later to be erroneous. I have described some of the questions which make this episode an  example of the various contradictory factors which influence the  knowledge process of nature at an individual, collective, sociological and philosophical  level.

I have described various types of contradictions which are at work in the quest for truths about nature and its processes. Errors are an intrinsic part of this quest, as a temporary result of the network of these contradictions. They may, after some time, be corrected. Invalidating what was thought a crucial result may result in invalidating  a whole theoretical framework. Or it may result in a minor rearrangement of the theory, at the cost of abandoning a minor detail, leaving intact its main architecture. 

Scientific reports cannot be believed at face value before they are submitted to critical assessments, to various verification processes. The latter combine new experimental and theoretical efforts by multiple actors. Free confrontations of opinions and experimental results through scientific journals, colloquia, international conferences, etc., are an essential, vital component for the accumulation of undisputable truths. The latter are the result of a social cristallization of true beliefs, which no further improvements of experimental techniques or theory developments 

	\section{Appendix}\label{append}
	 In all superconducting materials known until 1986, the specific heat was well known to vanish exponentially with $T$. This was easily explained within the electron pairing mechanism in singlets described in 1958 in ref.\cite{BCS}. The latter mechanism creates a forbidden energy gap $\Delta \simeq k_bT_c$ at the Fermi surface, which freezes electronic excitations for temperatures $T < T_C$,  where $T_c$ is the temperature below which superconductivity appears. 
	
	In the initial  RVB proposal,  a fermionic Fermi surface of pure spin excitations coexisted with superconductivity down to $T=0$, so that electronic spin excitations where just as possible as electronic charge excitations in a normal metal. As a result, the low temperature specific heat had to vary with $T$ as in a normal metal, i.e. $C_v \propto T$.
	
	The RVB proposal started  with the known fact that the undoped $CuO$ atomic layers carries one localized electron on each crystal site and is an insulator, almost a textbook example of a magnetic insulator. This localization of electrons on crystal sites is the result of the strong repulsion $U$, compared to the bandwidth $W$   : expression (\ref{II})  then is the correct starting point. Conventional wisdom had it in the seventies that the ground state of $CuO$ atomic layers would be a 2D insulating antiferromagnet, with electronic spins  up alternating from atom to atom with down spins.
	
	When two atoms labelled $i$ and $j$ in a crystal carry each an unpaired localized  electron, the quantum state for the two electrons may have different forms, depending on the overlap between the atomic wave functions. Roughly speaking, the two spins 1/2 may couple to form a total spin $S=1$,  or they may (approximately) couple in alternate directions  :spin projection $S_{z,i}= +1/2$ on one atom, and $S_{z,j}=-1/2$  on the other. If there is an ordered crystalline array of magnetic atoms, the two possibilities lead either to ferromagnetism (with all spins along the same direction) or to antiferromagnetism with an alternate order of spins along two opposite directions.
	
	But there is a third possibility  : both spins $1/2$ at sites $i$ and $j$  may combine in a superposition of amplitudes to form a state $S=0$ The superposition is that of state $|+ ->$ with $|-+>$, i.e. the state with spin $0$ which is a singlet  : $1/\sqrt{2}(|+-> - |-+>)$ sharing the two atoms $i$ and $j$. A singlet is a boson, formed with two spin 1/2 particles, which are fermions.
	
	The Resonating Valence Bond proposal, first introduced in 1973 by Phil Anderson about a 1D model of localized spins \cite{PWA2} (the "railroad threstle"), is that the ground state of an ordered 2D crystal, such as $CuO$ 2D atomic layer,  with one unpaired electron spin per atom is the superposition of tensor products of all possible singlet states coupling electrons in the crystal in  singlets associating electrons on atoms two by two. (In fact the ground state of a 2D crystal with localized electrons on crystal sites was studied in the late  eighties  by Monte Carlo computations which did not support the RVB ground state proposal; they found an antiferromagnetic ground state).
	
 Superconductivity, according to the RVB scheme,  was the result of doping the magnetic insulator with holes (i.e. of suppressing electrons by chemical doping) in a RVB ground state. 

A doped hole in the RVB ground state breaks a singlet, thus liberates an unpaired spin 1/2, and an atomic site with zero localized spin  : a spinless hole. Then each new particle, the spinless hole on one hand, and the neutral spin 1/2 can migrate in the crystal because of the atomic wave functions overlap.

A stunning result of this analysis is that, within the RVB scheme,  in a strongly correlated 2D crystal, doping of holes results in a separation of charge and spin. In a weakly interacting electronic liquid (such as a normal metal like $Cu$,  each electron carries simultaneously charge and spin 1/2.
 
At finite doping, the conclusion is that there exist a population of  charged bosons called holons, and a population of neutral fermions with spin 1/2, called spinons. A liquid of bosons condenses in a a superfluid state (such as $^4He$).  Charged holons may condense in a charged  superfluid state  : a superconductor. A liquid of free spinons obeys Fermi statistics, and has a Fermi surface of spin excitations, similar to that of electrons in a normal metal.

With his revolutionary RVB proposal, Phil Anderson in 1986 seemed to take into account the main original properties of $LBCO$, and to provide an explanation for the superconductivity in a doped magnetic insulator. 

The other starting point of the theory, i. e. expression (\ref{I}), which was at the basis of my first paper on HTS, leads to a different analysis of the ground state of the undoped crystal  : with one electron per atom in extended wave functions, the crystal has a half filled band of delocalized electronic states (the filled band has twice as many states because of the degeneracy due to the  electron spin). However, in the undoped 2D crystal, the Fermi surface is a square. In that case (a special case of so-called "nesting Fermi Surface"), weak interactions cause an instability of the normal metallic state and give rise to a Spin Density Wave (SDW) which has alternate spin density direction on each atom. This SDW causes an energy  gap at the Fermi Surface, and thus the ground state is a (weak) antiferromagnetic insulator. Schulz \cite{schulz} showed that in that case, there is a competition between the insulating SDW ground state and superconductivity. I showed with my students \cite{ledererlmp} in 1986 that a more realistic Fermi surface allows superconductivity to dominate. 

Given the similarity between the results of the  two scenarii, it is eventually not surprising if  a correct theory combines concepts of both.

\end{document}